\documentclass[pre,floatfix,twocolumn,showpacs,a4]{revtex4}
\usepackage{graphicx}
\usepackage{dcolumn}
\usepackage{amsmath}
\usepackage{amssymb}
\usepackage{tabularx}
\usepackage{epsfig}
\usepackage{mathrsfs}
\usepackage{rotating}

\graphicspath{{./}}

\begin{document}
\title[Short Title]{Network Automata: Coupling structure and function in real-world networks}
\author{David M.D. Smith$^{1,2,3,4}$}
\email{d.smith3@physics.ox.ac.uk}
\author{Jukka-Pekka Onnela$^{2,3,5}$}
\author{Chiu Fan Lee$^{2}$}
\author{Mark Fricker$^{6}$}
\author{Neil F. Johnson$^{7}$}
\affiliation{$^1$Centre for Mathematical Biology, Oxford University, Oxford OX1 3LB, U.K.}
\affiliation{$^2$Department of Physics, Clarendon Laboratory, Oxford University, Oxford OX1 3PU, U.K.}
\affiliation{$^3$CABDyN Comlexity Centre, Oxford University, Oxford, OX1 1HP, U.K.}
\affiliation{$^4$Oxford Centre for Integrative Systems Biology, Department of Biochemistry, Oxford University, South Parks Road, Oxford, OX1 3QU, U.K.}
\affiliation{$^5$DBEC, Helsinki, University of Technology, P.O. Box 9203, FIN-02015 HUT, Finland.}
%\affiliation{$^1$Sa\"id Business School, Oxford University, Oxford OX1 1HP, U.K.}
%\affiliation{$^2$Physics Department, Clarendon Laboratory, University of Oxford, Oxford, OX1 3PU, U.K.}
%\affiliation{$^3$Laboratory of Computational Engineering, Helsinki University of Technology, FIN-02015HUT, Finland}
\affiliation{$^6$Department of Plant Sciences, University of Oxford, Oxford, OX1 3RB, U.K.}
\affiliation{$^7$Physics Department, University of Miami, Coral Gables, Florida FL 33124, USA.}

\date{\today}
\begin{abstract}
We introduce Network Automata, a framework which couples the topological evolution of a network to its structure. It is useful for dealing with networks in which the topology evolves according to some specified microscopic rules and, simultaneously, there is a dynamic process taking place on the network that both depends on its structure but is also capable of modifying it. It is a generic framework for modeling systems in which network structure, dynamics, and function are interrelated. At the practical level, this framework allows for easy implementation of the microscopic rules involved in such systems. To demonstrate the approach, we develop a class of simple biologically inspired models of fungal growth. 
\end{abstract}

\pacs{89.75.Fb, 87.15.A-, 87.85.Xd}
% JP:  ----------------------------------------------------------------------------------------------------
% PACS (Physics and Astronomy Classification Scheme):
%87. 	Biological and medical physics
%87.17.-d 	Cellular structure and processes
%87.17.Aa 	Theory and modeling; computer simulation
%87.18.-h 	Multicellular phenomena
%87.18.Bb 	Computer simulation
%89. 	Other areas of applied and interdisciplinary physics
%89.75.Fb 	Structures and organization in complex systems
%89.75.Hc 	Networks and genealogical trees
%87.80.Vt 	Dynamical, regulatory, and integrative biology
% JP:  ----------------------------------------------------------------------------------------------------
\maketitle

\section{Introduction}
\label{sec:intro}

The network framework has proved very successful in the study of various complex interacting systems. In the network
description, the interacting elements are depicted as nodes and the interactions between the elements are represented by links connecting the corresponding nodes. The science of complex networks has progressed very quickly in the last few years, and some excellent reviews have been written covering both the methodology and key results \cite{ABReview,NewmanReview,DorogovtsevReview}. The strength of the complex network paradigm lies in 
its ability to capture some of the essential structural characteristics of interacting systems while reducing the details of 
both the elements and their interactions. Consequently, the early complex network literature was almost exclusively focused on 
structural properties of networks. 

%\begin{figure}
%\begin{center}
%\includegraphics[width=7.0cm]{schemetest.eps}
%\caption{ \label{fig:schematic} (Color online) The functional, stochastic and restricted behaviors which can be encompassed by the FDN and NA %frameworks. Also illustrated is the position of the Barab\'asi-Albert (BA) network growth model, the random attachment (RA) network growth model, %the Watts-Strogatz small world model (WS), the Game of Life cellular automata (GOL), the biologically inspired models (BIM) introduced in %Section~\ref{sec:biol} and the Self-Organised Criticality model of Fronczak, Fronczak and Holyst (FFH)~\cite{SOCN1}.}
%\end{center} 
%\end{figure}

In many out-of-equilibrium growing networks, the evolution at a given time is dependent on the configuration of the network at that time, as exemplified in the preferential attachment model of Barab\'asi and Albert~\cite{Barabasi}. In this paper, we develop the Network Automata (NA) framework, which can be seen as a natural extension of the Cellular Automata (CA) framework \cite{Wolfram}. When employed to model dynamic, functional networks, the NA framework not only removes ambiguity at an implementation level through exhaustive specification of the microscopic ruleset employed, it also provides a platform for comparison between apparently different network algorithms.

%We describe different variants of NA in Sections \ref{sec:NA} and \ref{sec:restricted} which are illustrated schematically in %Fig.~\ref{fig:schematic}. To demonstrate the versatility of NA itself, we show how some familiar network models can be recast in the NA framework %in Section \ref{sec:stochastic}. The NA framework not only removes ambiguity at an implementation level through exhaustive specification of the %microscopic ruleset employed, it also provides a platform for comparison between apparently different network algorithms.

%Consider a situation in which the topology of a network evolves while there is simultaneously some process taking place on it. 

While structural properties remain important in constraining the behavior of the system, there 
is also significant interest in understanding dynamical processes taking place on networks \cite{MotterReview}. Indeed, the marriage between structure and dynamic processes is so strong that the behavior of dynamic processes can be used to detect structure~\cite{Laplacian}. While a network's topology constrains the type of dynamics that may unfold on it, in many scenarios the dynamical process may influence the subsequent evolution of the topology - meaning that the structural properties of the network are coupled to its function. A real-world example might be the evolution of transport links within a city. The dynamics of the human population using this network in turn affect the reinforcement or removal of the those transport links and the feedback process is apparent. A similar situation may arise in the context of social networks, where one's current social 
opportunities and dynamics are limited by the existing network structure, but they can be widened by extending the network. 

%-- see, for example, the specific Self-Organised Criticality model of Fronczak, Fronczak and Holyst~\cite{SOCN1} and the model of Zimmerman {\it et al}~\cite{Zimmerman}.

\begin{figure}[!ht]
\begin{center}
\includegraphics[width=0.4\textwidth]{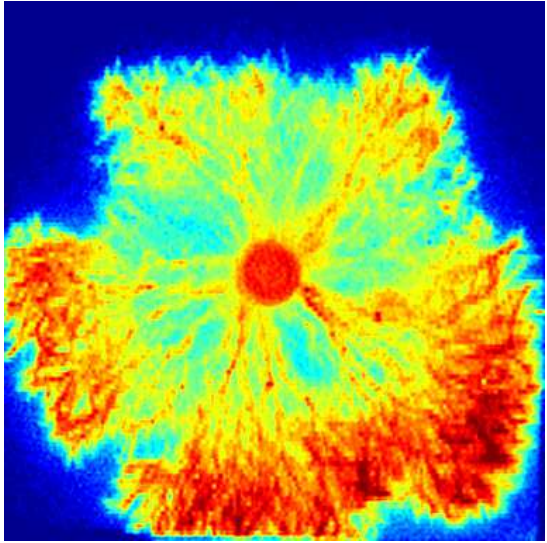}
\caption{ \label{fungi} (Color online) Transport of non-metabolised, radioactively labelled amino acid within 
\emph{Phanerochaete velutina}, a foraging woodland fungi. The image was taken using photon counting scintillation imaging~\cite{Tlalka2003,Tlalka2002}.}
\end{center} 
\end{figure}

There have been several specific attempts in the literature to inter-relate a network's structure, dynamics, and function \cite{SOCN1, Zimmerman, Gross, Bornholdt}. However, the present paper distinguishes itself by providing a {\em generic} framework for dealing with these types of systems. While many network generating algorithms and Cellular Automata models can be easily reproduced within the Network Automata framework~\cite{Thesis}, its usefulness is exemplified when it is applied to systems in which topological evolution is coupled to a dynamic process occurring upon the network. Its contribution to the network field lies in the fact that it enables a {\em precise} specification of the microscopic rules underlying the structural \emph{and} functional evolution of a given network-based system, thereby enabling comparison between different classes of network. We describe the framework in Section~\ref{sec:NA} and as  a demonstration, we introduce a class of progressively more realistic, biologically-inspired models of woodland fungal growth in Section~\ref{sec:bio}. Although the models consist of simple rulesets, they are nevertheless capable of producing structures qualitatively similar to mycelial network development (e.g. Fig.~\ref{fungi}). This example is \emph{Phanerochaete velutina}, a foraging saprotrophic woodland fungus that builds an adaptive network to translocate resource~\cite{Tlalka2002,Tlalka2003,Bebber2007}. The transport flux is oscillatory and still not fully understood~\cite{Fricker2008, Tlalka2007,Tlalka2002,Tlalka2003}.

%We expect there to be many application domains for Network Automata from social to biological systems.

\section{Network Automata}\label{sec:NA}
Consider an arbitrary weighted or unweighted, directed or undirected network at some time $t$ which is to be grown to some size $N_{tot}$.
As nodes might be added to the system at each time step, the network would conventionally be considered to be growing. An alternative representation is to treat the system as being of size $N_{tot}$ at all times where at some time $t$ many of the nodes have no links. Information regarding the network's topology  is entirely encompassed within the adjacency matrix $\mathbf{A}(t)$ which is of dimension $N_{tot}\times N_{tot}$. The matrix holds information about which links exist, their direction and, perhaps, weights. One might consider the evolution of the network as a process that alters the elements within this adjacency matrix, updating the attributes of any of the possible links which could exist in the system. We base a framework of network growth around this concept. If the microscopic ruleset governing the network's evolution is solely related to quantities which can be derived from the network's current topology (and hence from $\mathbf{A}(t)$) then the evolution of the network might be expressed in terms of some operation $F$ acting upon the adjacency matrix:
\begin{eqnarray}
\mathbf{A}(t+1)&=&F\big(\mathbf{A}(t)\big).
\end{eqnarray}
The ruleset employed could relate to any property of the nodes (their degree, betweenness, clustering and so on) or the links (weights or direction) and might be deterministic or stochastic.

%~\footnote{The concept 
%of a Network Automaton has been alluded to before although not formally defined \cite{Wolfram}.}

This concept can be achieved in practice by visiting \emph{all} possible links within the adjacency matrix
$\mathbf{A}(t)$  for a network comprising $N_{tot}$ nodes (whether part of a component or not) once every time step. The update as to the nature of the link at the next time step (its existence or its weight or direction) is then prescribed by the ruleset~\footnote{Because each link can see the requisite information at the start of the time step, the system is synchronously updated. This implementation could then encompasses asynchronous updates through application over longer timescales.}. The ruleset can take the form of a lookup table in which the \emph{state} of a link is evaluated and the \emph{color} (existence/direction/weight) is prescribed for the next timestep. This update process is analogous to the update of a cell within a conventional Cellular Automaton~\cite{CA} \emph{except} that it acts upon the connectivity (or neighbourhood) of an node, thereby generating $\mathbf{A}(t+1)$. This link-orientated update is a generic description of a dynamic network and all the essential features of that network's evolution are then contained within the exhaustive ruleset. We describe such a system as a \emph{Network Automaton}~\footnote{The concept of a Network Automaton has been alluded to before although not formally defined \cite{Wolfram}.}. No restriction has yet been made as to the directionality or weight of links. 

Now consider a situation in which the topology of a network evolves while there is simultaneously some process taking place on the 
network. At any given time the topology of the network constrains the type of dynamics that may unfold on it. However, the 
dynamical process may influence the subdequent topological evolution of the network, so that its structural properties are coupled to its 
function and vice versa.  The ruleset governing the topological update process relates not only to network related quantities but also functional aspects of the nodes and/or links. Since the functional process requires a network on which to perform, we decouple the evolution of the network into two distinct phases, namely, that affecting its topology and that governing the functional process. Writing the functional information (relating to nodes and/or links) at some time $t$ as some matrix $\mathbf{S}(t)$, the formal description of the evolution can be expressed in terms of some operations $F$ and $G$ as 
\begin{eqnarray}
\mathbf{A}(t+1)&=&F\big(\mathbf{A}(t),\mathbf{S}(t)\big),\label{eqn:definition}\\
\mathbf{S}(t+1)&=&G\big(\mathbf{A}(t+1),\mathbf{S}(t)\big). \label{eqn:definition2}
\end{eqnarray}
This expression states that the network evolves according to some process, which is determined by its own current topology $\mathbf{A}(t)$, \emph{and also} by some attributes of its nodes and links that includes function-based information $\mathbf{S}(t)$~\footnote{The discrete time recurrence relation is more appropriate than the continuous time coupled differential equation analogy $\frac{d\mathbf{A}}{dt}=f(\mathbf{A},\mathbf{S})$, $\frac{d\mathbf{S}}{dt}=g(\mathbf{A},\mathbf{S})$ in that it allows discontinuous changes in the system's update.}. The functional process then occurs on this network to generate the new set of information $\mathbf{S}(t+1)$. The global state of the system is encompassed by the matrices $\mathbf{A}$ and $\mathbf{S}$.
%The global state of the automaton is thus determined by $\mathbf{A}(t)$ and $\mathbf{S}(t)$.

%growth of the fungi and distribution of resources within it
%This is a test~\ref{eqn:definition} and ~\ref{eqn:definition2} and 
\section{Biologically Inspired Model}\label{sec:bio}
We now construct a series of simple models of woodland fungi~\cite{Tlalka2003, Tlalka2008,Bebber2007,Fricker2008} to demonstrate the versatility of the Network Automata framework. 
Although the model, which describes the growth of the fungi and distribution of resources within it, is biologically inspired, its aim is not to incorporate a large number biological details. Instead, we adopt a minimalist approach to emulate fungal growth and internal nutrient transport from a small set of microscopic rules. We start by specifying biologically naive and mathematically simple update rules (Ruleset {\bf a}) which govern the topological part of the system evolution as in Eq.~\ref{eqn:definition}. This is paired to a simple functional update process (Process {\bf 1}) representing Eq.~\ref{eqn:definition2}. The two combined fully define the model.  We then modify both the topological rules (Ruleset {\bf b} and {\bf c}) and functional process (Process {\bf 2}) according to some basic physical and biological considerations providing $6$ distinct models. The end product may serve as a platform for more elaborate future models of fungal growth, demonstrating the effectiveness of using the framework.

\begin{figure}[!ht]
\begin{center}
\includegraphics[width=8.6cm]{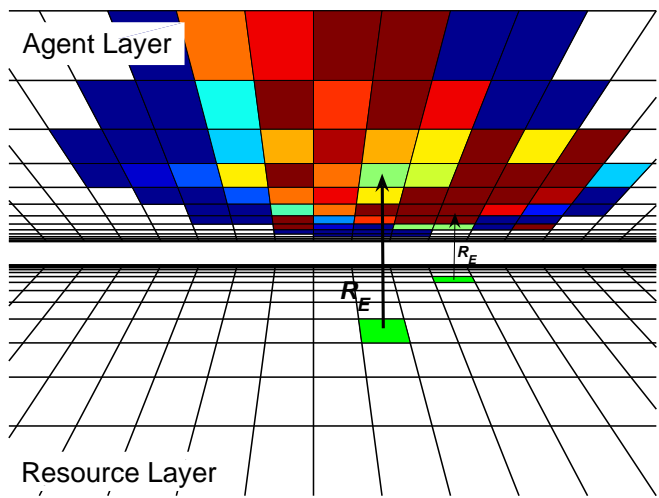}
\caption{ \label{fig:resourceagents} (Color online) A biologically inspired multi-agent model whereby the agent layer is 
superimposed upon a resource layer. Agents above a resource can accumulate resources at some rate $R_E$.}
\end{center} 
\end{figure}

Consider a system of agents who might each be interpreted as a cell in a two dimensional lattice. The connectivity between agents 
is North, South, East and West reflecting a possible connectivity of $4$, so that agents are restricted to local information 
as long range communication is assumed unlikely in the biological system. The agent layer is superimposed on a resource layer 
as in Fig.~\ref{fig:resourceagents}. The rules of the system are very simple. If an agent is above a resource, it absorbs that 
resource at some rate $R_E$.  The objective of each agent is two-fold: to grow into available space in the search for more resource, and to redistribute excess internal resource to support further exploration. We allow each agent to grow only one new neighbor (in a random direction) at a particular time, but only if the agent has resources to do so. To mimic active transport of resources to the growing tip, an agent passes resources to 
its neighbor cell provided that the neighbor does not pass resources to it. We now endeavor to categorize this simple multi-agent 
system in the NA framework. This serves not only to clarify any ambiguities that arise in the programming of a multi-agent system,
 but also as a potential aid to improving efficiency in that the required iteration and information storage/retrieval aspects are clearly defined by the rulesets imposed. 
\begin{table*}
\begin{center}
\begin{tabular}{|c|c|c|c||c|c||c|c||c|c|}
\multicolumn{4}{c}{} & \multicolumn{2}{c}{Model {\bf a}} & \multicolumn{2}{c}{Model {\bf b}} 
& \multicolumn{2}{c}{Model {\bf c}}\\
\hline
\multicolumn{4}{|c||}{time $t$} & \multicolumn{2}{c||}{ time $t+1$} & \multicolumn{2}{c||}{time $t+1$} 
& \multicolumn{2}{c|}{ time $t+1$}\\
\hline
 $A_{i,j}(t)$ & $A_{j,i}(t)$ & $\phi(S_i(t))$ & $\phi(S_j(t))$ & $A_{i,j}(t+1)$ & $A_{j,i}(t+1)$ & $A_{i,j}(t+1)$ & $A_{j,i}(t+1)$ 
& $A_{i,j}(t+1)$ & $A_{j,i}(t+1)$\\
\hline
0&0&0&0&0&0&0&0&0&0\\
0&0&0&1&0	& $\omega\left(\frac{1}{d-k_j(t)}\right)$ & 0	& $\omega\left(\frac{1}{d-k_j(t)}\right)$ & 0	& 
$\omega\left(\frac{g}{d-k_j(t)}\right)$\\
0&0&1&0&$\omega\left(\frac{1}{d-k_i(t)}\right)$&0 & $\omega\left(\frac{1}{d-k_i(t)}\right)$ & 0 & 
$\omega\left(\frac{g}{d-k_i(t)}\right) $ &0\\
0&0&1&1&$\omega\left(\frac{1}{2}\right)$&$1-A_{i,j}(t+1)$ & $\omega\left(\frac{1}{2}\right)$ & $1-A_{i,j}(t+1)$ & 
$\omega\left(\frac{1}{2}\right)$ & $1-A_{i,j}(t+1)$\\
0&1&0&0&0&1 & 0 & 1 & 0 & 1\\
0&1&0&1&0&1 & 0 &1 & 0 & 1\\
0&1&1&0&0&1 & 0 &1 & 0 & 1\\
0&1&1&1&0&1 & $\delta(_{{\rm in}}k_{i},d)\omega\left(\frac{1}{d}\right)$ & $1-A_{i,j}(t+1)$ & 
$\delta(_{{\rm in}}k_{i},d)\omega\left(\frac{1}{d}\right)$ & $1-A_{i,j}(t+1)$\\
1&0&0&0&1&0&1&0&1&0\\
1&0&0&1&1&0&1&0&1&0\\
1&0&1&0&1&0&1&0&1&0\\
1&0&1&1&1&0&$1-A_{j,i}(t+1)$ & $\delta(_{{\rm in}}k_{i},d)\omega\left(\frac{1}{d}\right)$ & $1-A_{j,i}(t+1)$ & 
$\delta(_{{\rm in}}k_{i},d)\omega\left(\frac{1}{d}\right)$\\
\hline
\end{tabular}
\caption{\label{tab:bio} Different rulesets (the columns labeled `time $t+1$') for three biologically inspired models: ({\bf a}) the 
simplest scenario, ({\bf b}) incorporating conservation of resources, and ({\bf c}) implementing a delay factor (see text for details). As $A_{i,j}$ and $A_{j,i}$ are mutually exclusive, there are only $12$ possible states of a link at time $t$.}
\end{center}
\end{table*}	

Let us first look at the growth (topological) stage. Each cell or agent represents a node and the boundary between two cells through which resource is passed is represented by a link. Consider that the information upon which the topological ruleset will act to update the attributes of a link in the network is simply the amount of resource that each of the two nodes has at each end of the link and their in and out-degrees. We can simply write the functional information as a vector such that $S_i(t)$ refers to  the resource (functional variable) that agent (node) $i$ has at time $t$. For clarity, we can express some topological information as vectors too, such that the element $k_i(t)$ represents the total degree of a node, $_{{\rm out}}k_{i}(t)$ its  out-degree which, in turn, provides its in-degree $_{{\rm in}}k_{i}(t)$, each of which is obtainable from the adjacency matrix $\mathbf{A}(t)$. 
%The information available to the both the functional and topological update rulesets are thus vectors of dimension $N_{tot}$ which is the total %number of possible cells in the system.

We will grow this Network Automaton in an unweighted but directed adjacency matrix $\mathbf{A}$ so that if $A_{i,j}=1$ the link  
exists and is directed from $i$ to $j$, whereas if $A_{j,i}=1$ the link exists and is directed from $j$ to $i$. 
If neither  $A_{i,j}=1$ nor $A_{j,i}=1$ then the link does not exist. 
Here $A_{i,j}$ and $A_{j,i}$ are mutually exclusive. As each node has limited possible connectivity (here $4$), we only consider the subset of links in the system which could possibly exist. The structural update process runs through all of these possible links and each pair of nodes which could be connected is considered once. The link attributes $A_{i,j}$ and $A_{j,i}$ are then updated at the same time. 

We can now write the network update procedure for this topological update as Ruleset {\bf a} in terms of an exhaustive truth table as in Table \ref{tab:bio}. Note that if  $A_{i,j}(t)=A_{j,i}(t)=0$, the total degree of both nodes $i$ and $j$ is less than the connectivity $d$ of the lattice in which the system evolves. We denote the outcome of a Bernoulli trial as $\omega(p)$ such that $P(\omega(p)=1)=p$ and conversely $P(\omega(p)=0)=1-p$. We employ a step function $\phi(x)$ defined as $\phi(x)=1$ for $x>0$ and $\phi(x)=0$ for $x\le0$.
%Also, the dependence on the underlying lattice is implicit and, as such, its cumbersome presence will not be explicitly included in the ruleset. 

We now describe the resource distribution (functional) stage and start by mapping the adjacency matrix 
$\mathbf{A}(t+1)$ to a normalised transition matrix $\mathbf{T}(t+1)$ describing the flow of resource between adjacent cells:
\begin{eqnarray}\label{eqn:transitiona}
T_{i,j}(t+1)&=&\left\{\begin{array}{l l}
 A_{i,j}(t+1)/{_{{\rm out}}k_i(t+1)} &\textrm{for $_{{\rm out}}k_{i}(t+1)>0$}\\
 0&\textrm{for $_{{\rm out}}k_i(t+1)=0$}
 \end{array}\right.\nonumber\\
 T_{i,i}(t+1)&=&\left\{\begin{array}{l l}
 0 &\textrm{for $_{{\rm out}}k_{i}(t+1)>0$}\\
 1&\textrm{for $_{{\rm out}}k_{i}(t+1)=0$.}
 \end{array}\right.
\end{eqnarray}
Through this process, an agent distributes resource equally amongst those neighbours which are not transmitting resource to it. 

We can then write the update for the resource distribution process as
\begin{eqnarray}\label{eqn:transitionb}
\underline{S}(t+1) &=& \mathbf{T}^{\dag}(t+1) \underline{S}(t) +\underline{\xi}(t), 
\end{eqnarray}
where the vector $\underline{\xi}$ corresponds to the accumulation of resource by agents from the resource (substrate) layer. We impose the constraint that only ``alive" (i.e. active) agents can accumulate resource through this process so
\begin{eqnarray}\label{eqn:transitionc}
\xi_i(t) &=& R_E ~ \phi(S_i(t)) ~ L_i ,
\end{eqnarray}
where the vector $\underline{L}$ denotes the (binary) existence of resource at position of node (agent) $i$ in the resource layer, $R_E$ is the rate at which an agent accumulates the resource and $\phi(x)$ is the step function defined earlier.  This resource accumulation from the substrate could be made time dependent (i.e. finite resources) although here we will not consider this effect. Equations ~\ref{eqn:transitiona}, ~\ref{eqn:transitionb} and ~\ref{eqn:transitionc} constitute Process {\bf 1} and represent the functional update stage of Eq.~\ref{eqn:definition2}.

\begin{figure}[!htb]
\begin{center}
\includegraphics[width=4.2cm]{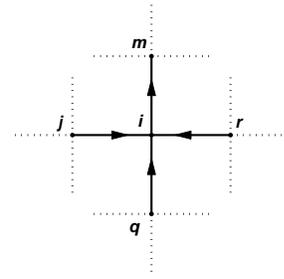}
\caption{ \label{fig:nodeinflux} The influx of resource into node labelled $i$. The amount this node receives from node $j$ is  
related to the out-degree of node $j$. This amount would be expressed as $S_j(t)/_{{\rm out}}k_{j}(t+1)$.}
\end{center} 
\end{figure}

For the example of Fig.~\ref{fig:nodeinflux}, the amount of resources that node $i$ has at time $t+1$ is
\begin{eqnarray}
S_i(t+1) & = & S_j(t)\frac{A_{j,i}(t+1)}{_{{\rm out}}k_{j}(t+1)} + S_m(t)\frac{A_{m,i}(t+1)}{_{{\rm out}}k_{m}(t+1)}\nonumber\\
 & & + S_r(t)\frac{A_{r,i}(t+1)}{_{{\rm out}}k_{r}(t+1)} + S_q(t)\frac{A_{q,i}(t+1)}{_{{\rm out}}k_{q}(t+1)}\nonumber\\
 &=&S_j(t)T_{j,i}(t+1)~~~+~~~S_r(t)T_{r,i}(t+1)\nonumber\\
 {}&{}&{~~~~~~~~~~~+~~S_q(t)T_{q,i}(t+1).}	
% & = &\frac{S_j(t)}{k_{o,j}(t+1)}+\frac{S_r(t)}{k_{o,r}(t+1)}+\frac{S_q(t)}{k_{o,q}(t+1)}.
\end{eqnarray}

\begin{figure}[!ht]
\begin{center}
\includegraphics[width=4.2cm]{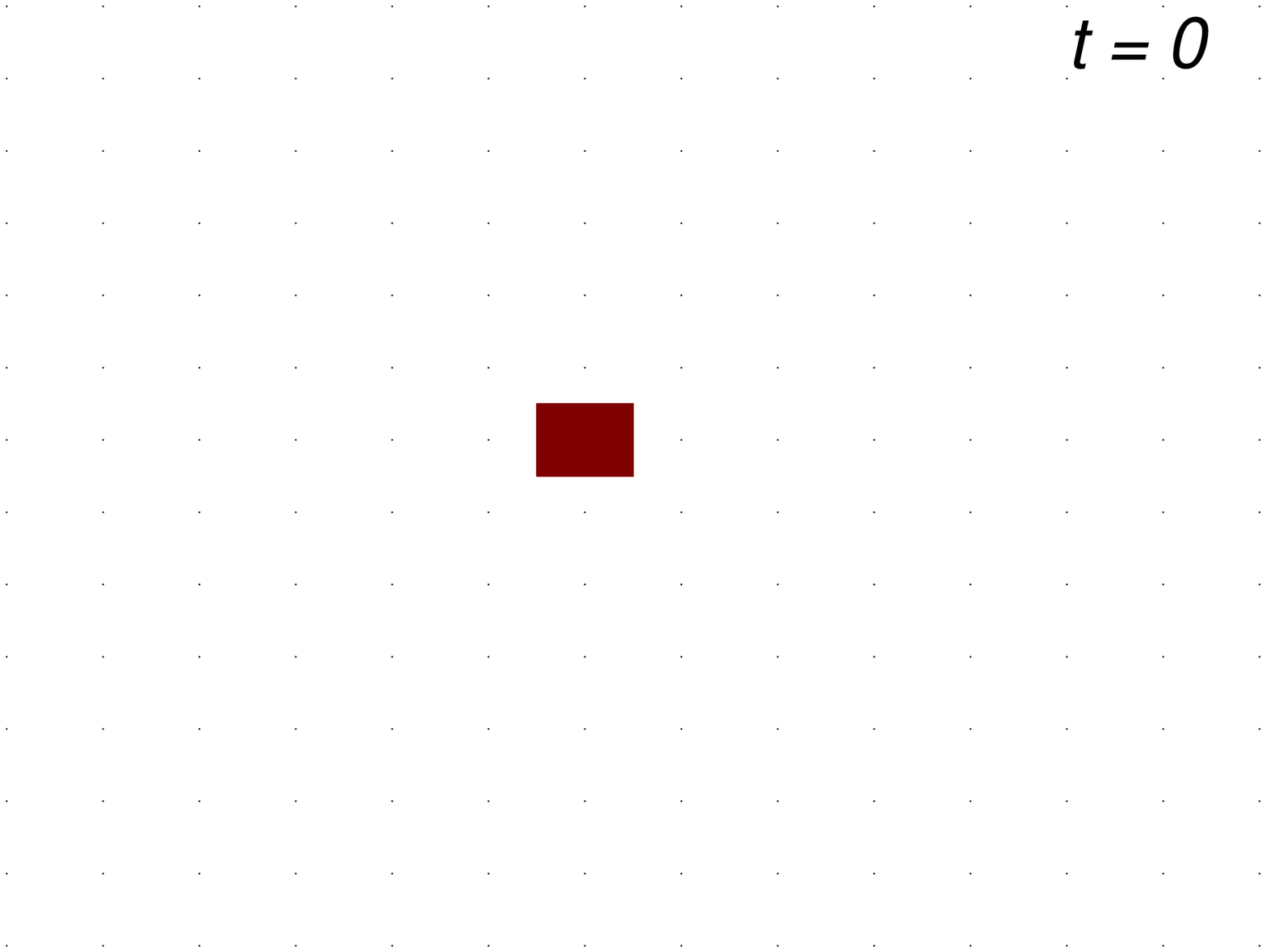}
\includegraphics[width=4.2cm]{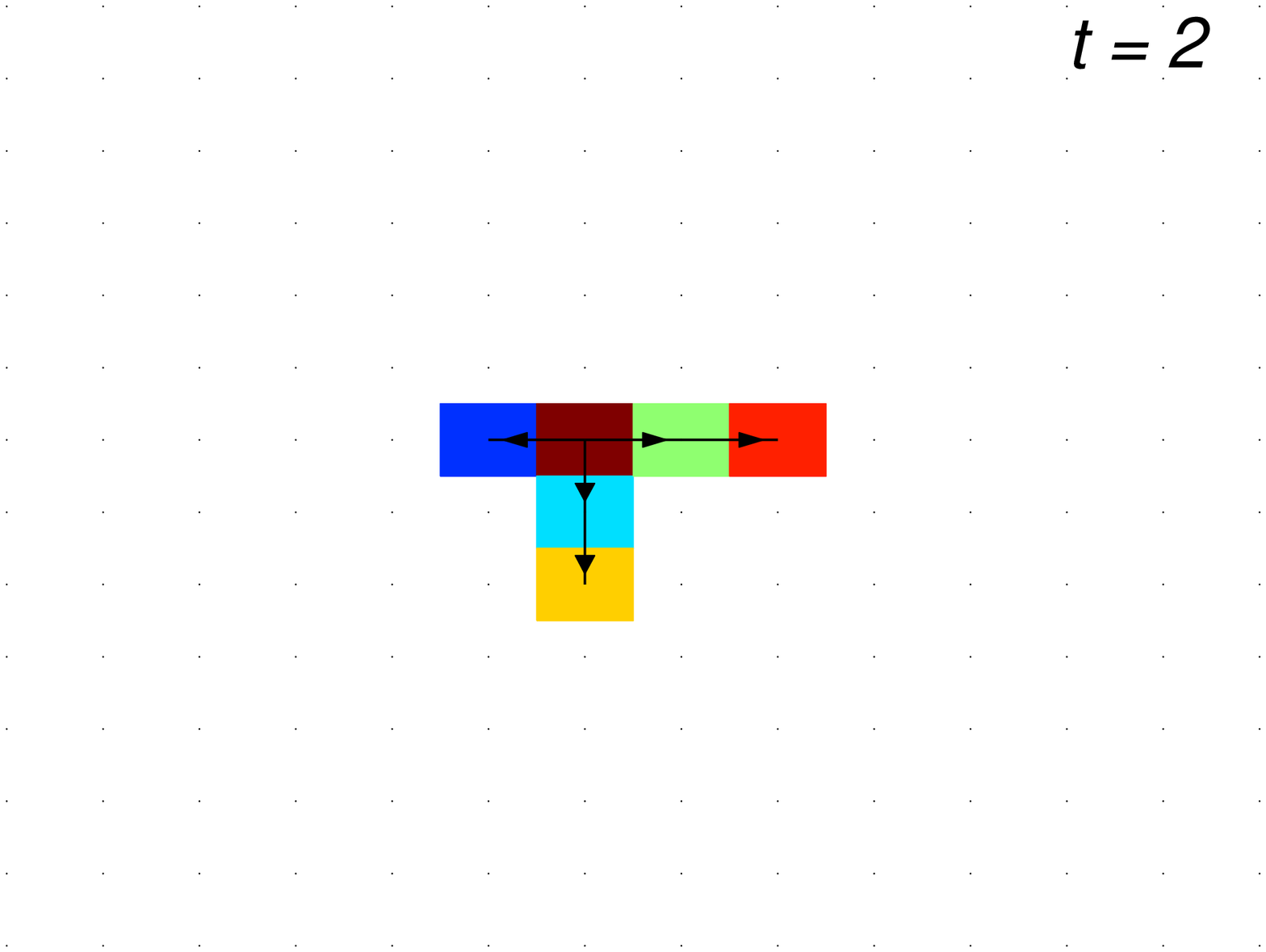}
\includegraphics[width=4.2cm]{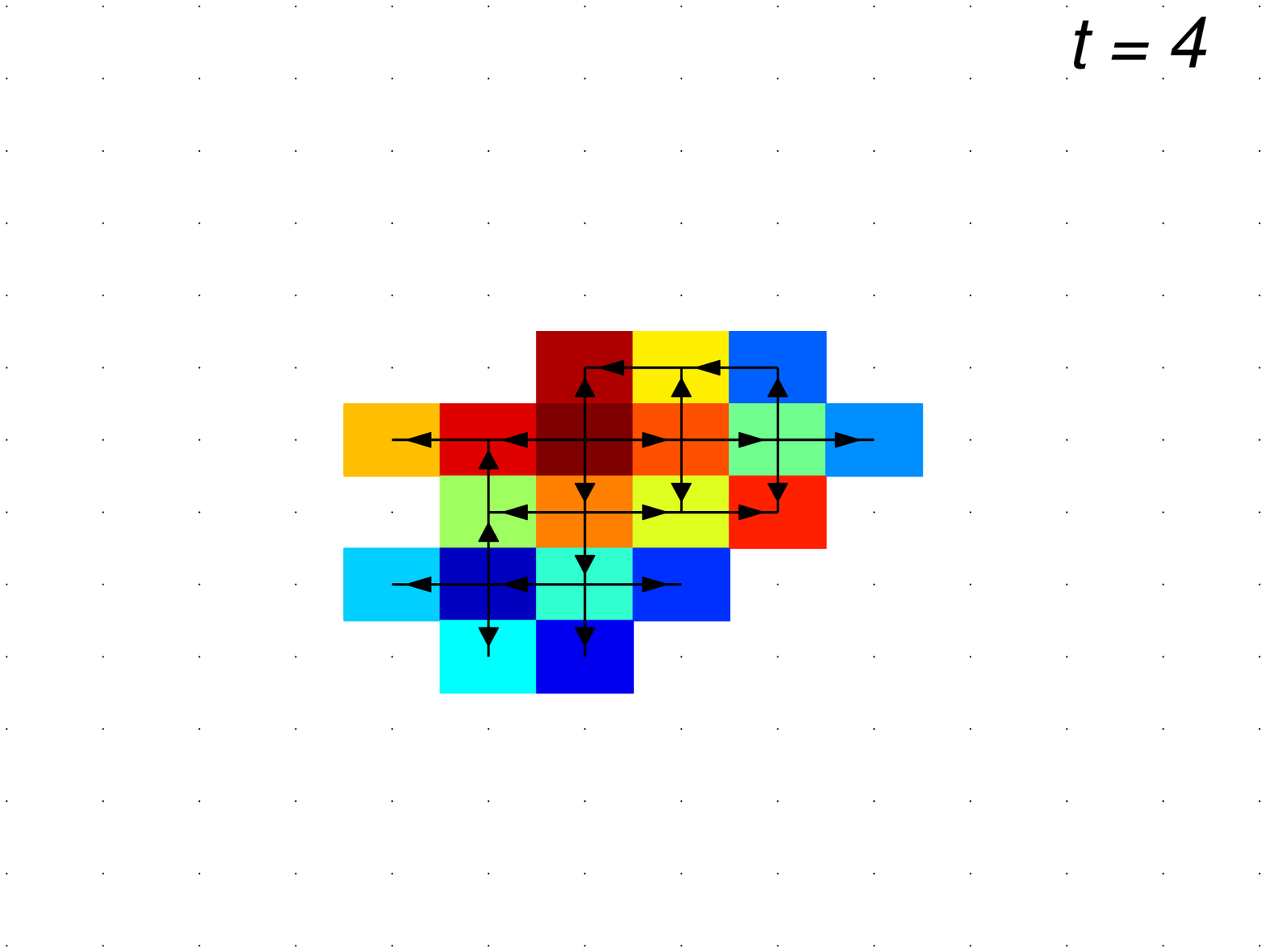}
\includegraphics[width=4.2cm]{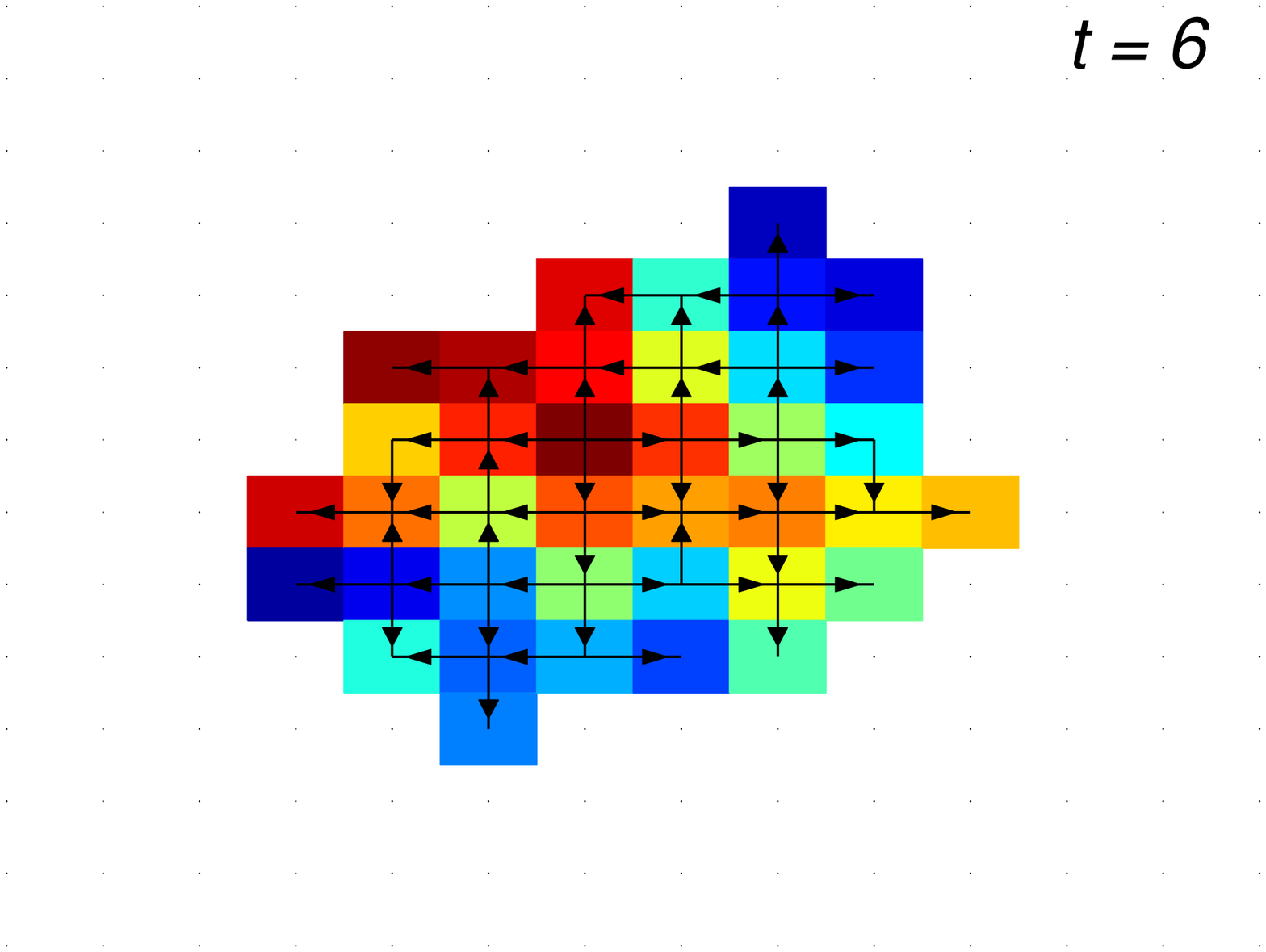}
\caption{ \label{fig:evolution} (Color online) The evolution of the biologically inspired NA over $6$ timesteps. The colors represent the amount of  
resource $S_i(t)$ a node has (the functional aspect) superimposed on a directed network (the structural aspect).}
\end{center} 
%\end{figure}
%\begin{figure}[!htbp]
%\begin{center}
%\includegraphics[width=7.0cm]{Fungus3.eps}
%\caption{ \label{fig:square} (Color online) Incorporating conservation and consumption of resources. The simulation is seeded with  
%one resource, one agent, $R_E=20,000$, $R_C=1$ and after $500$ time steps there are $15,288$ agents. The simulation used the ruleset  
%of Model {\bf b} of Table \ref{tab:bio}, and the functional update is that of Eqs.~\ref{eqn:transition2},~\ref{eqn:explicit1},  
%\ref{eqn:eatrate}.}
%\end{center} 
%\end{figure}
\end{figure}

\begin{figure*}[!htbp]
\begin{center}
\includegraphics[width=0.32\textwidth]{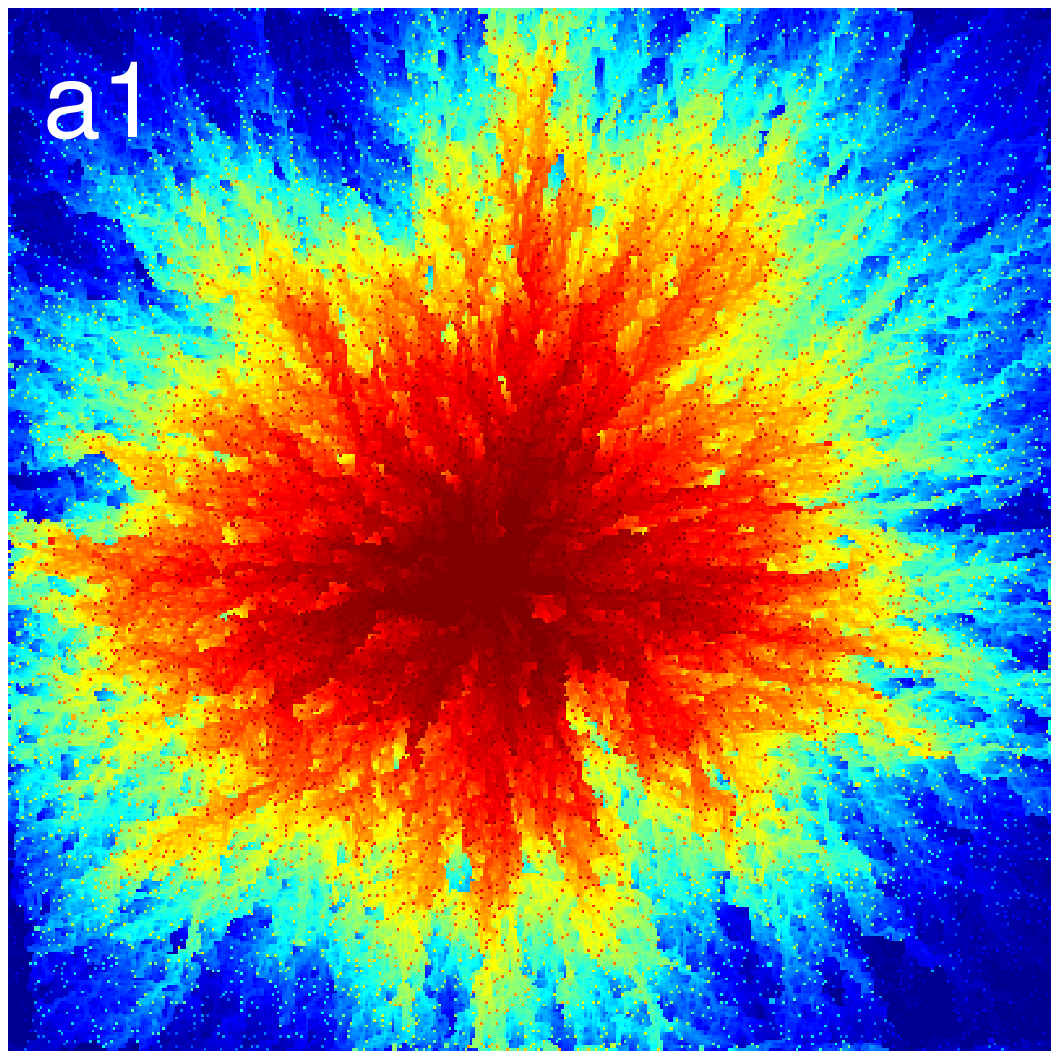}
\includegraphics[width=0.32\textwidth]{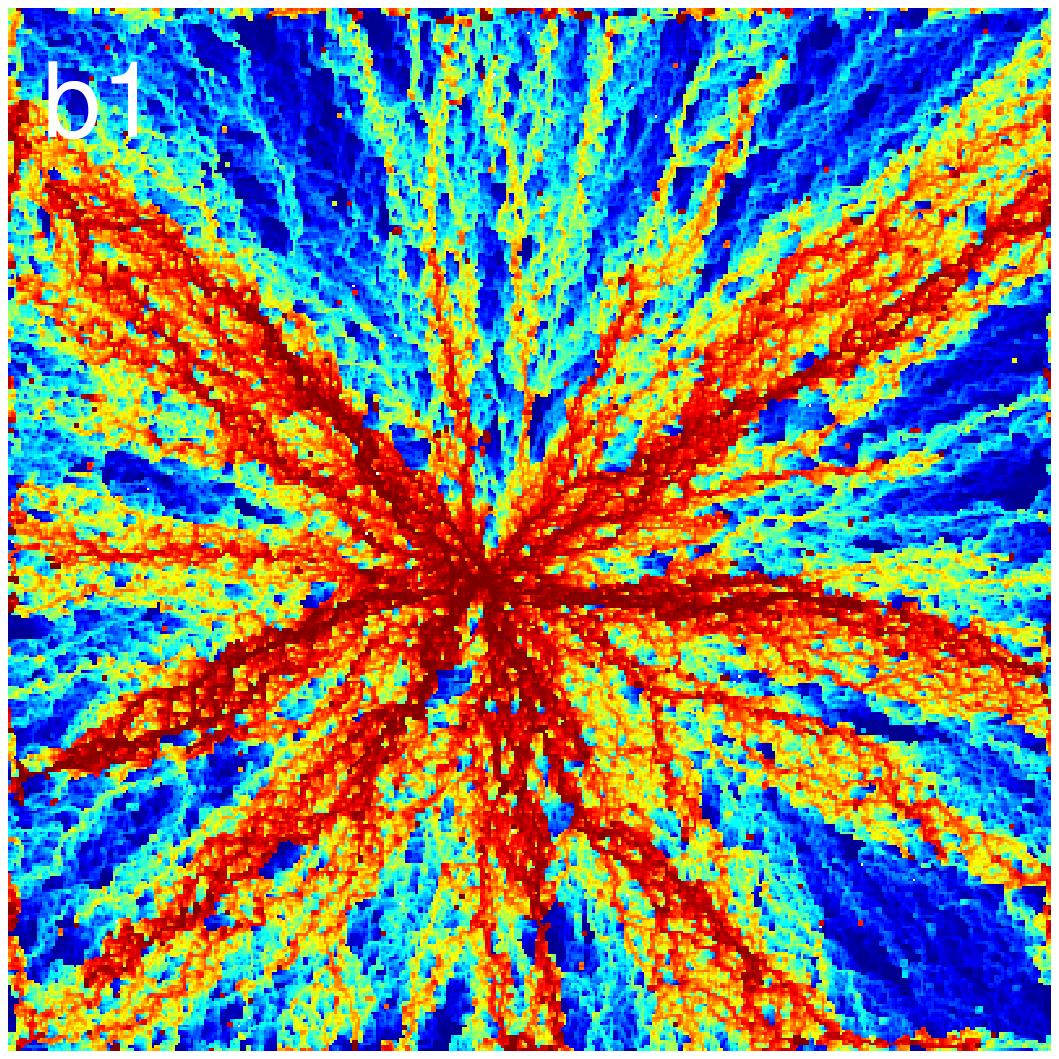}
\includegraphics[width=0.32\textwidth]{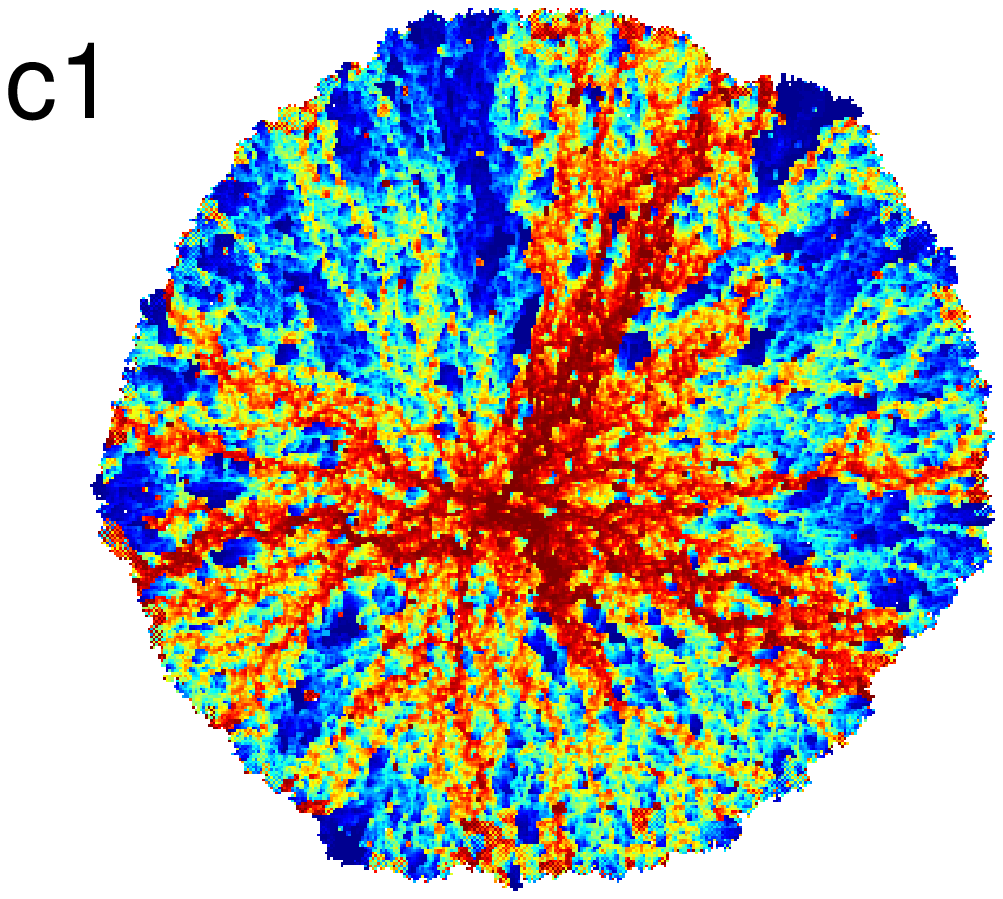}
\includegraphics[width=0.32\textwidth]{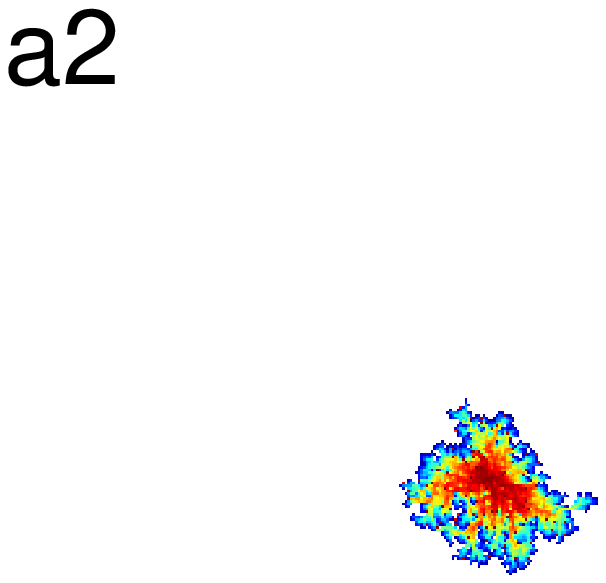}
\includegraphics[width=0.32\textwidth]{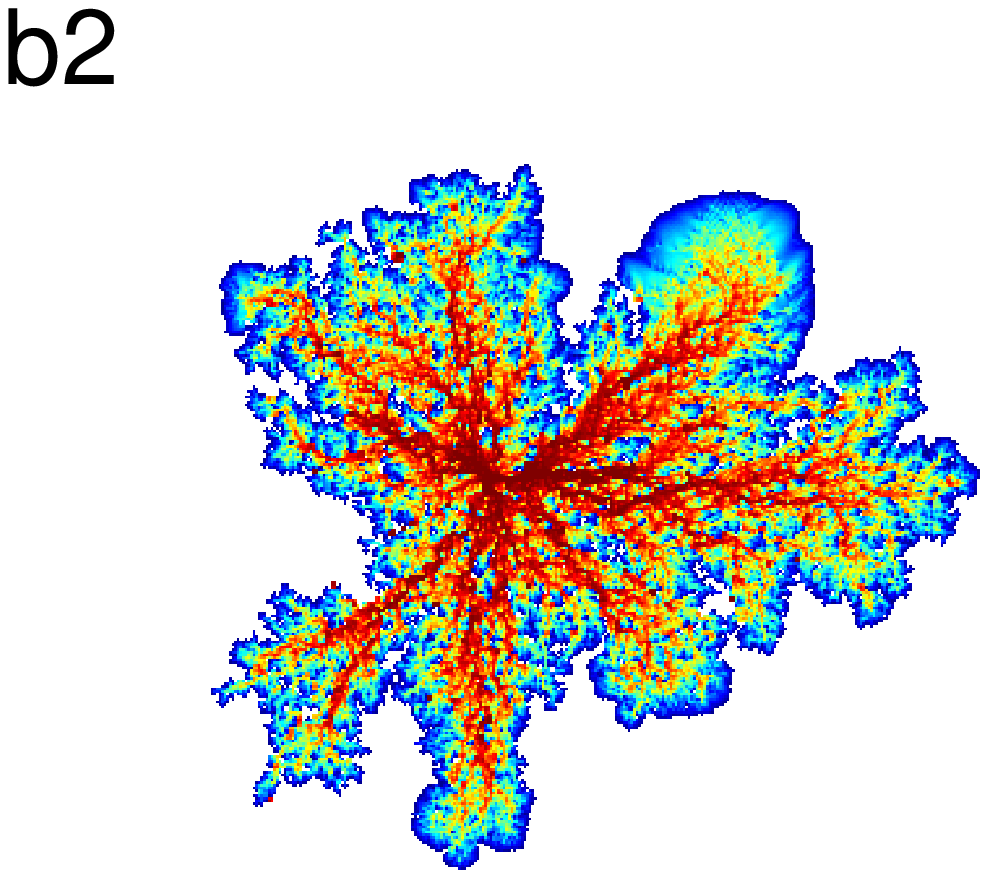}
\includegraphics[width=0.32\textwidth]{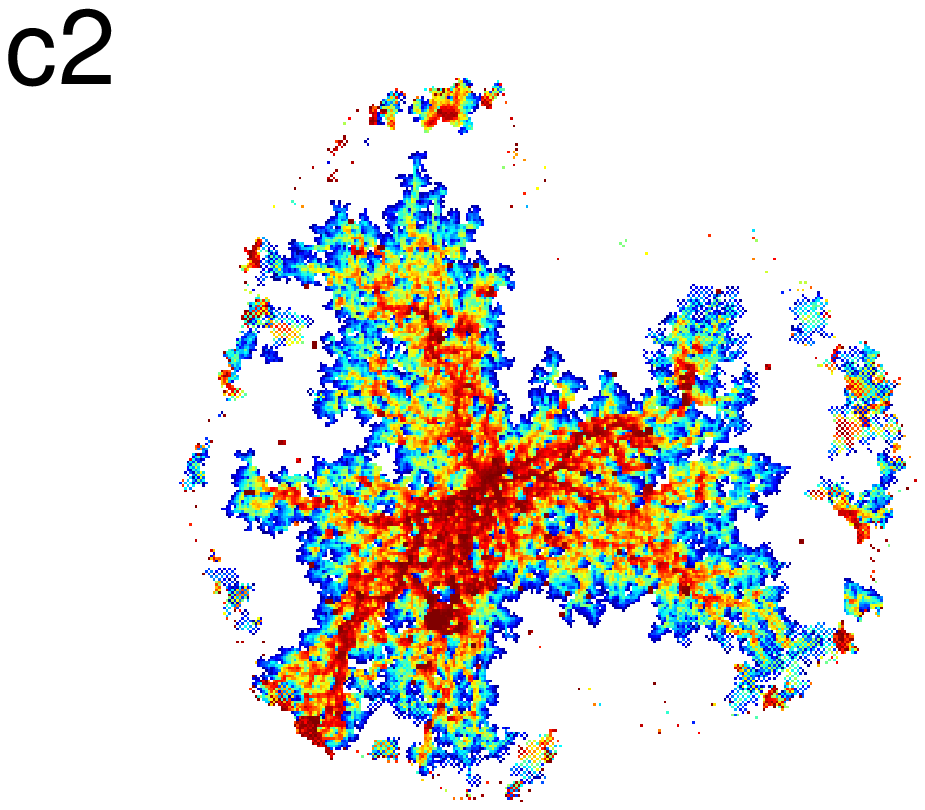}
\caption{ \label{fig:all} (Color online) Simulated output for the Rulesets {\bf a}, {\bf b} and {\bf c} and functional updates of Processes {\bf 1} and {\bf 2}. For all models, the lattice is of dimension $400\times 400$ with periodic boundary conditions. Each was seeded with a single initial agent above a single (infinite) resource. The rate of resource accumulation for that agent is $R_E=80,000$ and the simulation was run for $2000$ time steps for each model. For Process {\bf 2} the residual consumption rate of each agent is set at $R_C=1$. For Ruleset {\bf c}, the delay factor is $g=0.1$.}
\end{center} 
\end{figure*}

We can now observe the Network Automaton in operation as shown in Fig.~\ref{fig:evolution}. We start with a single node $\eta$ above a  
single food source so that at time $t=0$ the agent has some resource.  In the initial configuration the adjacency matrix is all  
zeros $A_{i,j}(0) = 0 ~\forall i,j$, and the resource information vector is all zeros except $S_{\eta}(0)=R_E$ such that the  
initial agent has amount $R_E$.  For this example, the resource accumulation vector $\underline{\xi}(t)$ is also all zeros except  
$\xi_{\eta}(t) = R_E \, \forall \, t$. We observe both the network and functional aspect of the system. The nodes (agents) are  
superimposed on the directed network, and the amount of resources a node has, is indicated by its color  ranging from blue (low concentration) to dark red (high concentration)~\cite{Thesis,Fricker2008}. Only nodes that have resources are included and  
the result is independent of the choice of $R_E$. A longer simulation is shown in Fig.~\ref{fig:all}.

Note that under this ruleset and functional update stage an agent might accumulate resource indefinitely. By allowing a node which has in-degree  equal to the maximum connectivity ($_{{\rm in}}k_{i}=d$) to randomly flip a direction of one of its links overcomes this biologically unfeasible behavior and results in the topological update of (Ruleset {\bf b}) where the Kronecker delta function is defined as $\delta(x,y) = 0$ for $x\ne y$ and $\delta(x,y) = 1$ for $x = y$. A simulation of this ruleset coupled with the functional update of Process {\bf 1} is shown in Fig.~\ref{fig:all} and emergent canalized flux channels are clear. Such channels have been observed experimentally in a wide class  of real biological fungi~\cite{Tlalka2003}.


In Rulesets {\bf a} and {\bf b}, the physical transport of nutrient is of comparable speed to that of the growth, which is clearly not  reasonable for most biological systems. A further development can be made to incorporate two different time scales in the model so that the growth and redistribution of resources take place at different rates, leading to Ruleset {\bf c}. To establish this feature within in the model, we can delay the growth process by  introducing a parameter $g$ to the stochastic growth terms in the ruleset, so that an agent grows on average one neighbor every  $1/g$ time steps. This is depicted explicitly in Table \ref{tab:bio}. The result of such a delay is the build up of resource towards the periphery of the object as one might expect~\cite{Tlalka2003}. This is illustrated in Fig.~\ref{fungi} for the real fungi and in Fig.~\ref{fig:all} for the model.

We can also incorporate consumption of resources by agents in the model by making further modifications to the functional update stage,  
resulting in Process {\bf 2}. If an agent has more than some residual consumption amount, $R_C$, then this rate of consumption
is deducted. If the agent has less than this value, all of that agent's resources are removed and that agent might be  
considered dead. The initial stage in the functional update process can be described in a similar manner to Eq.~\ref{eqn:transitionb}:
\begin{eqnarray}\label{eqn:explicit1}
\underline{S}'(t+1) &=& \mathbf{T}^{\dag}(t+1) \underline{S}(t) + \underline{\xi}(t), 
\end{eqnarray} 
where the resource distribution is described by the transition matrix $\mathbf{T}(t+1)$ defined in Eq.~\ref{eqn:transitiona} and the accumulation of resource from the substrate is described by the vector $\underline{\xi}(t)$ defined in Eq.~\ref{eqn:transitionc}. We can then write the update for $\underline{S}(t+1)$, including consumption of resource as
\begin{eqnarray}\label{eqn:eatrate}
S_i(t+1) & = &\phi(S'_i(t+1) - R_C)(S'_i(t+1)-R_C), \nonumber \\ 
\end{eqnarray}
which makes use of the step function $\phi(x)$. Only agents (nodes) active in the network can  
accumulate resources from the resource layer. The effect of this ``cost" of living clearly limits the potential size of the system. Equations ~\ref{eqn:transitiona}, ~\ref{eqn:transitionc}, ~\ref{eqn:explicit1} and ~\ref{eqn:eatrate} constitute Process {\bf 2} and the effect of using this functional update are illustrated in Fig.~\ref{fig:all}.

%The ruleset of each progressively more detailed model is based on that of its predecessor as seen in Table~\ref{tab:bio}.
\section{Concluding remarks and discussion}
\label{sec:disc}
In this paper we have developed the concepts of Network Automata (NA) which can couple evolution and function of complex networks by using simple microscopic rules at the level of nodes and links. We have demonstrated the practicability of the framework by applying it to a class of biologically inspired models, which produce qualitatively similar canalised flow patterns to those observed in real woodland fungi~\cite{Tlalka2003, Tlalka2007,Tlalka2008,Fricker2008}. This suggests that for organisms that have to adapt their morphology to a variable environment, function may play a crucial role in determining structure. The well defined and simple rulesets not only make replication straightforward but also aid implementation at the programming level. In the study of emergent phenomena involving networks, this  
framework enables concise and clear establishment of microscopic rules. One can easily think of more complex rulesets to more  
accurately model a real system, such as adding transport costs or finite resources, both of which can easily be accomplished, or even a time-dependent ruleset. It is then interesting to  
ask what level of complexity is required to more accurately model real biological systems. 

We expect there to be many application domains for Network Automata from social to biological systems. One can envisage applying the NA framework also to discrete differential equation modeling and Diffusion Limited Aggregation (DLA) systems and, indeed, to any system in which the dynamics of network topology is related to the function performed thereon. There are certain biological systems which can quasi-solve increasingly complex problems in a constant time. An example is \emph{Physarum polycephalum}, the true slime mold, which approximately identifies the Steiner
points when placed upon multiple food sources~	\cite{Toshi}. Given that it might be possible to model these systems within the NA framework, it might suggest how to design a hardware based implementation to perform similar calculations in constant time. It is interesting to
pose the question as to what kind of problems could be solved by such a system and how complex the microscopic rules would be for a given problem. This would reflect the minimum length of ruleset that would have to be employed by the system in both the network and
functional update stages.

D.M.D.S. acknowledges funding by EPSRC and BBSRC and J.-P.O. by a Wolfson College
Junior Research Fellowship (Oxford). D.M.D.S acknowledges funding from the European Union (MMCOMNET)
for part of this research. We thank Felix Reed-Tsochas for comments and suggestions.
%
%\appendix
%\section{Conservation and consumption of resources}
%\label{sec:con}
%We can incorporate conservation of resources in the model by modifying the update stage of the functional process. This requires  
%rewriting the transition matrix as
%\begin{eqnarray}\label{23}
%T_{t+1}(i,j)&=&\left\{\begin{array}{l l}
% \frac{A_{t+1}(i,j)}{k_{o,i}(t+1)} &\textrm{for $k_{o,i}(t+1)>0$}\\
% 0&\textrm{for $k_{o,i}(t+1)=0$}
% \end{array}\right.\nonumber\\
% T_{t+1}(i,i)&=&\left\{\begin{array}{l l}
% 0 &\textrm{for $k_{o,i}(t+1)>0$}\\
% 1&\textrm{for $k_{o,i}(t+1)=0$.}
% \end{array}\right.
%\end{eqnarray}
%Note that the above modification allows an agent to accumulate resource indefinitely if its in-degree is equal to the degree of  
%the underlying lattice, i.e. $k_i(t) = d$. This undesirable feature can be overcome by having this agent flip the direction of,  
%on average, one of its $d$ links. To implement this we make use of the Kronecker delta function defined as
%\begin{eqnarray}\label{eqn:delta}
%\delta(x,y)&=&\left\{\begin{array}{ll}
%0&\textrm{for $x\ne y$}\\
%1&\textrm{for $x = y$.}\end{array}\right.
%\end{eqnarray}

\end{document}